\documentclass[11pt]{article}
\usepackage[utf8]{inputenc}
\usepackage[a4paper,margin=24mm]{geometry}
\usepackage{graphicx}
\usepackage{amsmath}
\usepackage{booktabs}
\usepackage{tabularx}
\usepackage{array}
\usepackage{url}
\DeclareUrlCommand\hash{}
\usepackage[numbers,sort&compress,super]{natbib}
\usepackage{caption}
\newcolumntype{P}[1]{>{\raggedright\arraybackslash}p{#1}}
\newcommand{\U}{\mathrm{U}}
\newcommand{\run}[1]{\texttt{#1}}

\title{A multi-fidelity dataset of two-dimensional steady laminar channel flow around obstacles}
\author{Alper Saydam\textsuperscript{1,*} and Bar\i\c{s} Akg\"un\textsuperscript{1}\\[6pt]
  \small \textsuperscript{1}KUIS AI Center, Ko\c{c} University, Istanbul, Turkey\\
  \small \textsuperscript{*}Corresponding author: asaydam21@ku.edu.tr}
\date{}

\begin{document}
\maketitle

\begin{abstract}
\noindent Surrogates that correct inexpensive flow simulations need the same flows solved at several fidelities and an
estimate of the numerical error of the reference solutions. This dataset contains 4,500 two-dimensional, steady,
laminar channel flows past a single obstacle. Each is solved in five numerical configurations: second-order convection
on meshes with cell sizes of 2, 1, 0.5 and 0.25\,mm, and first-order convection on the 1\,mm mesh. A seeded generator
draws the obstacles from five shape families in three size classes. All 22,500 OpenFOAM runs were iterated to initial
residuals of $10^{-6}$, and each run's residuals, iterations and provenance are recorded. Every run provides the
velocity and pressure on a regular grid with geometry channels, and the pressure and wall shear along the obstacle.
Its 20 flow attributes, computed on the simulation mesh, include drag and lift split into pressure and friction parts.
The four second-order levels give an estimate of the discretisation uncertainty of 19 attributes of every case. Fixed
splits include three out-of-distribution test sets.
\end{abstract}

\section*{Background \& Summary}

Computational fluid dynamics (CFD) lets engineers test a design before they build a prototype. Every change to the
design needs a new simulation. A fine-mesh simulation can take hours, and a coarse-mesh one takes minutes but is less
accurate. Machine-learning surrogates for CFD predict a flow without a new simulation. They are trained on solved
flows, and many public datasets provide such
flows\cite{airfrans,drivaernet,flowbench,cfdbench,pdebench,thewell,drivaerml,unifoil,superwing}. A surrogate can also
correct a cheap solution, for example by predicting the fine-mesh flow or the drag from a coarse-mesh solution, and so
shorten the design cycle. Training such a surrogate needs the same flows solved at several fidelities. These flows
must cover geometries that vary enough for the surrogate to learn how the flow interacts with the obstacle. The
surrogate's accuracy can only be assessed with an estimate of the numerical error of its reference solutions, because
surrogates inherit the discretisation error of their training data\cite{jakeman}. In a study of wind-turbine wakes,
networks trained on data with a larger numerical uncertainty were less accurate\cite{ye}. If a surrogate is more
accurate than its labels, only a comparison with a finer reference can show it\cite{koehler}.

Among the datasets compared here (Table 1), most offer one resolution. In FlowBench\cite{flowbench}, the levels are
sub-sampled from one simulation. In a double-delta wing set, one level is computed with a cheaper vortex-lattice
model\cite{doubledelta}. MegaFlow2D, like this dataset, solves each of its two levels on its own mesh, but its flows
are transient\cite{megaflow}. The RANS study of Setinek et al.\cite{setinek} has two levels that differ in the
near-wall treatment. A later set from the same group has eleven meshes for each of 1,627 airfoils, and every run was
stopped after a fixed number of iterations\cite{setinek2026}. For chaotic turbulence, BLASTNet notes that matching
pairs of solved levels cannot be obtained, and it filters one solution instead\cite{blastnet}. Steady laminar flow has
no such limit, so each level can be a converged solution on its own mesh rather than a filtered copy of the finest
one.

The iterative solution converges slowly on fine meshes, so the stopping criterion decides how close a fine-mesh label
is to its steady solution. Where the other datasets give a criterion, it applies to the dataset as a whole (Table 1).
The criteria given are a residual target between $10^{-5}$ and $10^{-10}$ (sometimes together with steady forces), a
fixed number of iterations, or the convergence of the forces. In the double-delta set, 57 of 2,448 runs reached the
target only with a lower-order scheme\cite{doubledelta}. UniFoil and the double-delta set release the solver history
of every run, but none of the datasets compared shows how a looser criterion would change the labels. In a study
reported in Technical Validation, 113 cases were stopped at residuals of $10^{-3}$ (velocity) and $10^{-2}$
(pressure). When the runs start from rest, the 0.25\,mm labels lose about a third of their recirculation length. If
the coarse level also stops early, the coarse-to-fine difference is overstated by a factor of 36. When the runs start
from a coarse solution instead, the same criterion leaves the labels close to that solution, and the difference is
understated.

Mesh studies are usually made once, to choose the mesh\cite{megaflow,doubledelta}, or on a few reference
cases\cite{drivaernet,flowbench,superwing}. UniFoil judged a mesh study impractical for its half a million
runs\cite{unifoil}. The eleven-mesh airfoil set gives the error of every level against the finest one, which is taken
as the reference\cite{setinek2026}. None of the datasets compared estimates the numerical uncertainty of each label.
In a coarse-to-fine task, the error of a cheap solution and any improvement on it have to be read against the
uncertainty of the label used for comparison. Where the steady datasets compared give forces, these are totals. Only
the raw solver output of the double-delta set splits them into pressure and friction parts\cite{doubledelta}.

This dataset follows the setting of DeepCFD\cite{deepcfd}: two-dimensional steady laminar flow past one obstacle in a
120\,mm high channel. The inlet velocity and viscosity are the same, and the five obstacle families are similar.
DeepCFD offers fields only, at one resolution of about 1\,mm, and reports neither a stopping criterion nor a mesh
study.

MS-COFD-2DSL (Multi-Scale Central Object Flow Dataset, 2D Steady Laminar) provides 4,500 obstacle geometries in five
families and three size classes. The obstacles are rotated and shifted, and their blockage is limited. Each geometry
is solved to steady state in five numerical configurations (Table 2): with second-order convection on four meshes
with cell sizes of 2, 1, 0.5 and 0.25\,mm, and with first-order convection on the 1\,mm mesh. Each of the 22,500 runs
was iterated to initial residuals of $10^{-6}$, and its residuals, iterations, cost and provenance are released with
it. The four second-order levels form a refinement study of every case, from which the discretisation uncertainty of
19 scalar attributes is estimated. Each run provides the fields on a regular grid (180\,$\times$\,60 to
1440\,$\times$\,480 points) with geometry channels, and the pressure and wall shear along the obstacle. It also
provides 20 flow attributes computed on the simulation mesh. Fixed splits and three out-of-distribution test sets
(blockage, rotation, position) are part of the release, so that results can be compared. The solver setup was checked
against the DFG 2D-1 benchmark, and every run was audited. Runs that were processed or solved again reproduced the
released data. A separate repository defines tasks, metrics and baselines on these data and holds the uncertainty
estimates of every case\cite{code_bench}.

The data can be used for coarse-to-fine correction and super-resolution with paired, converged levels, and for
surrogate models of fields and of physical quantities. They also allow multi-fidelity learning across mesh size and
discretisation order, and studies of how the stopping criterion and the mesh affect learned models.

\begin{table}[t]
\centering\scriptsize
\caption{\textbf{This dataset and related steady or multi-resolution datasets.} Flows are steady unless marked
transient. Fidelity levels: number of levels and how they differ. Label uncertainty: estimate of the numerical error
of the labels, or the scope of the mesh study. Raw cases: whether the solver cases are released. RANS,
Reynolds-averaged Navier--Stokes; $C_D$, $C_L$, $C_M$, drag, lift and moment coefficients; $C_p$, $C_f$, pressure
and skin-friction coefficients; $p$, pressure; $\tau_w$, wall shear stress; RMS, root mean square; k, thousand;
history, solver history of every run released; n.r., not reported.}
\setlength{\tabcolsep}{3pt}
\begin{tabularx}{\textwidth}{@{}P{2.05cm}P{2.0cm}P{1.0cm}XXP{1.55cm}P{2.1cm}P{1.5cm}@{}}
\toprule
Dataset & Flow & Cases & Fidelity levels & Stopping criterion & Label uncertainty & Physical targets & Raw cases \\
\midrule
This dataset & 2D laminar, channel with one obstacle & 4,500 & 5 configurations: 4 meshes (2--0.25\,mm) and first
order at 1\,mm, all converged & per run, $<10^{-6}$, released and audited & estimated for 19 attributes per case & 20
on the simulation mesh, forces split; surface $p$, $\tau_w$ & all 22,500 runs \\
\midrule
DeepCFD\cite{deepcfd} & 2D laminar, channel with one obstacle & 981 & 1 & n.r. & n.r. & none & no \\
AirfRANS\cite{airfrans} & 2D RANS, airfoils & 1,000 & 1 & force convergence & n.r. & $C_D$, $C_L$, boundary layer &
yes \\
Setinek et al.\ 2025\cite{setinek} & 2D RANS, airfoils & 611 pairs & 2 (wall-resolved, wall functions) & n.r. & n.r.
& forces & n.r. \\
Setinek et al.\ 2026\cite{setinek2026} & 2D RANS, airfoils & 1,627 & 11 meshes (5k--212k nodes), solved
independently & fixed $2\times10^4$ or $4\times10^4$ iterations & n.r. (errors against the finest mesh) & $C_D$,
$C_L$ & no \\
UniFoil\cite{unifoil} & 2D RANS, airfoils, turbulent and transitional & ${>}$500k & 1 mesh; two turbulence models on
4,800 airfoils & n.r.; history & n.r. (mesh study ``impractical'') & $C_D$, $C_L$, surface fields & yes \\
SuperWing\cite{superwing} & 3D RANS, transonic wings & 28,856 & 1 & residual $<10^{-10}$, steady $C_L$ & reference
wings only & $C_D$, $C_L$, $C_M$, surface $C_p$, $C_f$ & no \\
Double-delta wings\cite{doubledelta} & 3D RANS, wings & 2,448 & 2 (vortex lattice, RANS) & continuity $10^{-6}$,
steady $C_D$; history & one geometry & $C_D$, $C_L$, $C_M$, split in raw files & solver output \\
DrivAerNet++\cite{drivaernet} & 3D RANS, cars & 8,000 & 1 & 7,000 iterations, RMS $<10^{-5}$ & one reference car &
$C_D$, $C_L$, surface $p$, $\tau_w$ & yes \\
FlowBench\cite{flowbench} (steady) & 2D/3D cavity with an object & $\approx$9,000 (2D) & 3, sub-sampled from one
solution & n.r. & reference cases only & $C_D$, $C_L$, Nusselt & no \\
MegaFlow2D\cite{megaflow} & 2D transient, circles, ellipses, a nozzle & 3,000 & 2, solved independently & --
(transient) & mesh chosen once & none & no (graph snapshots) \\
\bottomrule
\end{tabularx}
\end{table}

\section*{Methods}

\subsection*{Overview}
Each case is produced by one automated pipeline\cite{code_pipeline} (Fig.~1b). The obstacle is drawn from a seeded
random generator and meshed with gmsh. The flow is then solved to steady state with OpenFOAM in the five
configurations of Table 2, and each run is iterated until the initial residuals of velocity and pressure are below
$10^{-6}$. The converged fields are resampled to regular grids, and the flow attributes are computed from OpenFOAM
output on the simulation mesh. The complete case is then archived.

The 0.25\,mm run provides the labels. The 2\,mm run and the two 1\,mm runs are low-fidelity inputs.
The 2\,mm run is coarse but second order, whereas the first-order 1\,mm run is as fine as the second-order 1\,mm run
but more diffusive. The 0.5\,mm run is an intermediate refinement, so that the meshes can be compared in cost and
accuracy. The four second-order levels form a refinement study of every case, from which the discretisation
uncertainty of 19 scalar attributes is estimated. The two 1\,mm runs differ only in the convection scheme, so they
separate its effect from that of the mesh. Fidelity here means mesh resolution and convection order. The
configurations are not ordered by accuracy for every quantity. For example, the first-order 1\,mm run is closer to the
label than the 2\,mm run in pressure drag but further from it in total drag (Technical Validation). In what follows, a
configuration is one of the five numerical set-ups, a run is the solution of one configuration for one case, and the
levels are the four second-order meshes.

The runs have the same names as the folders of the release (Table 2). The 1\,mm mesh is the base mesh. On it,
\run{base\_lu} is solved with second-order linearUpwind convection and \run{base\_fo} with first-order upwind
convection. The runs \run{0.25x}, \run{4x} and \run{16x} use the other meshes. They are named after their number of
cells relative to the base mesh, which is about a quarter, four times and sixteen times as large.

\begin{figure}[t]
\centering
\includegraphics[width=\textwidth]{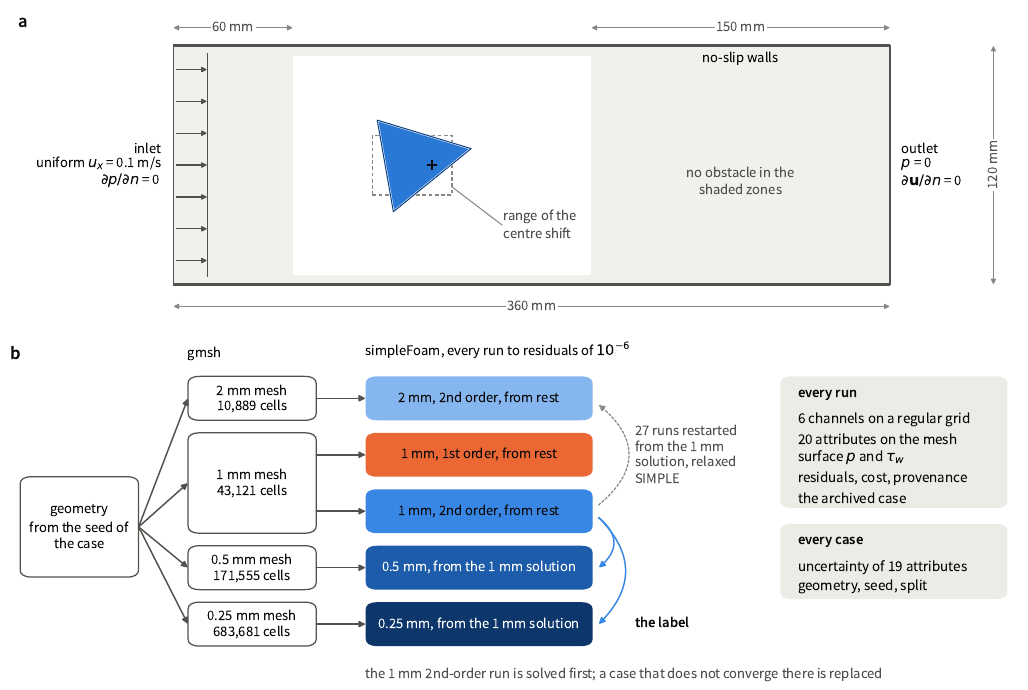}
\caption{\textbf{Flow configuration and the five runs of every case.} \textbf{a}, The channel, to scale, with its
boundary conditions. The obstacle keeps clear of the shaded zones, and its centre is shifted within the dashed range
around the nominal centre (+). The obstacle shown is the normal right triangle of Methods, Arrays. \textbf{b}, The meshes of a case
(median cell counts), its five runs and the fields from which each run starts.}
\end{figure}

\begin{table}[t]
\centering\footnotesize
\caption{\textbf{The five runs of every case.} Run: the folder name in the release. Cell counts, iterations and
solver times are medians over the 4,500 cases, with the 95th percentile of the iterations in brackets. Solver times
are for one core. \textsuperscript{a}The 27 relaxed 2\,mm runs started from the converged \run{base\_lu} solution
(Flow solution).}
\begin{tabular}{@{}llllrlrr@{}}
\toprule
Run & Cell size & Convection & Started from & Cells & Array grid & Iterations to $10^{-6}$ & Solver time \\
\midrule
\run{0.25x} & 2\,mm & second order & rest\textsuperscript{a} & 10,889 & 180\,$\times$\,60 & 209 (1,094) & 8\,s \\
\run{base\_fo} & 1\,mm & first order & rest & 43,121 & 360\,$\times$\,120 & 491 (1,426) & 77\,s \\
\run{base\_lu} & 1\,mm & second order & rest & 43,121 & 360\,$\times$\,120 & 587 (2,377) & 110\,s \\
\run{4x} & 0.5\,mm & second order & \run{base\_lu} & 171,555 & 720\,$\times$\,240 & 1,075 (3,501) & 19\,min \\
\run{16x} & 0.25\,mm & second order & \run{base\_lu} & 683,681 & 1440\,$\times$\,480 & 3,497 (11,109) & 5.6\,h \\
\bottomrule
\end{tabular}
\end{table}

\subsection*{Design choices}
The flow is steady and two-dimensional, so every configuration can be iterated to convergence on its own mesh,
and the four-level refinement could be afforded for 4,500 geometries. Even in two dimensions, a 0.25\,mm run takes
5.6 core-hours (median, Table 2). The resampled fields also lie on regular grids of the kind coarse-to-fine models
use. The inlet conditions are fixed, so the Reynolds number varies only through the obstacle height (Flow configuration).
With the inlet velocity and viscosity fixed, the effect of the geometry and of the numerical configuration is
isolated. At this inlet velocity, the steady solver converged up to the blockage limit (Geometry). The data do not
show how models transfer to other Reynolds numbers, to unsteady flow or to three dimensions.

The five shape families are modelled on those of DeepCFD. Rectangles, diamonds and triangles have sharp corners at
which the flow separates, and the spline ``circles'' have rounded outlines. The two triangle families, one pointing
upstream and one downstream, add an asymmetry. The rotation is limited to 45$^\circ$ to keep each family recognisable.
The mesh is refined uniformly because the uncertainty estimates of a case are based on four second-order levels of the
same type of mesh. As a result, the near-wall region is resolved no more finely than the rest of the domain. Locally
refined meshes are left to a later version. In a preliminary study of 30 geometries, the second-order 1\,mm solution
was already within 1--3\,\% of the mesh-independent value for almost every integral attribute. The 2\,mm level was
added because a coarse-to-fine task needs a coarser input than this. The 0.25\,mm level was the finest that could be
afforded for all cases (Computation and provenance).

The first-order 1\,mm run was cheap to add (77\,s per case, median). Attribute errors complement field errors, since a
field error can be small while the flow near the obstacle or in the wake is wrong. The pressure and friction parts of
the drag also show errors that cancel in the total. The out-of-distribution sets (Splits) test a mild extrapolation in
blockage, rotation or position. They contain no new shape families or flow conditions.

\subsection*{Flow configuration}
The domain is a channel of length $L = 360$\,mm and height $H = 120$\,mm, with the inlet at $x = 0$ (Fig.~1a). The
fluid is incompressible and Newtonian, with kinematic viscosity $\nu = 1\times10^{-4}$\,m$^2$\,s$^{-1}$. The steady
equations $\nabla\cdot\mathbf{u}=0$ and $(\mathbf{u}\cdot\nabla)\mathbf{u}=-\nabla p+\nu\nabla^2\mathbf{u}$ are solved
for the velocity $\mathbf{u}$ and the kinematic pressure $p$. At the inlet, the velocity is uniform,
$U_\mathrm{in} = 0.1$\,m\,s$^{-1}$, and the pressure has zero gradient. At the outlet, the velocity has zero gradient
and $p = 0$. On the channel walls and the obstacle, the velocity is zero (no-slip) and the pressure has zero gradient.
The Reynolds number is 120 based on the channel height, and $Re_h = U_\mathrm{in} h/\nu = 10$--48 based on the
obstacle height $h$, the height of its bounding box ($h/H$ is the blockage ratio). The inlet velocity is the same in
all cases, while the shape, size and placement of the obstacle vary (Geometry). OpenFOAM meshes are three-dimensional, so the mesh is one cell layer deep in $z$, normal to the
plane of the flow. The layer is 10\,mm thick, and its front and back faces are empty patches, which makes the
solution two-dimensional. Forces are reported for this depth. All fields are kinematic.

\subsection*{Geometry}
Five obstacle families are used (Fig.~2): rectangles, diamonds, left triangles, right triangles and ``circles''. The
apex of a left triangle points upstream, and that of a right triangle points downstream. The ``circles'' are closed
curves of four splines through perturbed points, which give a rounded but irregular shape. The obstacles come in three
size classes (smaller, normal and larger), with 300 cases for every family and class. Each case is drawn with its own
random generator, seeded with the string ``seed-size-family-index'' (global seed 2026), so that any case can be
regenerated on its own. Every mesh level of a case shares the same geometry.

A case is drawn in four steps. First, the width $W$ and the height are the nominal size of the family (rectangle and
diamond 50\,$\times$\,25\,mm, triangles 42.5\,$\times$\,42.5\,mm, circle 25\,$\times$\,25\,mm) times the scale of the
size class (0.65, 1 or 1.3). The width is further multiplied by $(1+u_w)$ and the height by $(1+u_h)$, with
$u_w \sim \mathcal{U}(-0.10, 0.10)$ and $u_h \sim \mathcal{U}(-0.20, 0.20)$. Second, the corner points are placed
around the nominal centre (130, 60)\,mm, and each is moved by a random offset of up to $0.17W$ ($0.14W$ for circles,
$0.085W$ for diamonds). A triangle is taken from such a perturbed rectangle, with the apex at the midpoint of its two
upstream or downstream corners. A circle is closed by four interpolating splines through the corners and one
perturbed point near the middle of each side. The exact construction is in the pipeline code\cite{code_pipeline}.
Third, the obstacle is rotated about its centroid by
$\theta \sim \mathcal{U}(-45^\circ, 45^\circ)$ and shifted by $\Delta x \sim \mathcal{U}(-30, 10)$\,mm and
$\Delta y \sim \mathcal{U}(-15, 15)$\,mm. The shift is biased upstream to leave room for the wake. Fourth, the obstacle
is drawn again if it comes closer than 5\,mm to a channel wall, 60\,mm to the inlet or 150\,mm to the outlet. It is
also drawn again if its outline crosses itself or if its blockage ratio (bounding-box height over $H$) exceeds 0.40.
The number of attempts is recorded.

The blockage limit was set in pilot runs of the production pipeline. Above a blockage of about 0.45, a growing share of obstacles did not
converge on the 1\,mm mesh, and their residuals stalled at $10^{-3}$--$10^{-2}$. At 0.40, almost every obstacle
converged. The stall may indicate the onset of unsteady flow in a confined wake, but this was not tested with
transient simulations. In the dataset, the blockage ratio ranges from 0.08 to 0.40 (median 0.28). A case whose 1\,mm
second-order run did not reach $10^{-6}$ within its iteration cap was dropped with all its runs. It was replaced by
the next reserve geometry of its size and family (indices 301--330, drawn from the same seed). No other run of a case
started before its 1\,mm second-order run had converged, so every case in the dataset has all five configurations.
Eighteen cases were dropped and replaced: 14 normal left triangles, 3 normal right triangles and 1 larger left
triangle. All had a blockage of 0.36--0.40 and largest residuals of $10^{-4}$ to $2\times10^{-2}$ at the
10,000-iteration cap. Dropping them removes part of the highest-blockage normal left triangles from the sampled
distribution. A dropped case may still have a steady solution, since other stalled runs converged once their
starting fields and solver relaxation were changed (see Flow solution). The dataset is therefore conditioned on
convergence under the production protocol.

\begin{figure}[t]
\centering
\includegraphics[width=\textwidth]{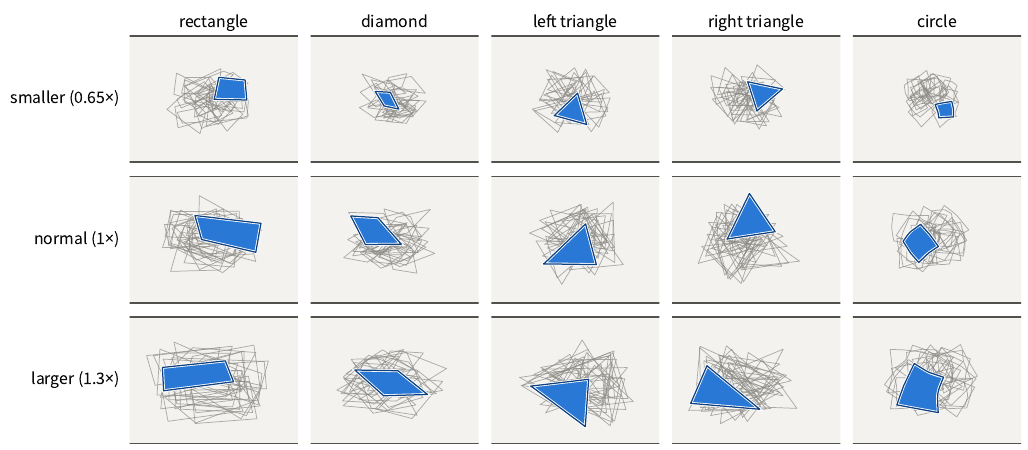}
\caption{\textbf{Obstacle families and size classes.} Columns show the five families and rows the three size
classes. Each panel shows the first 25 cases of its family and class as grey outlines, with the first one filled, on
the part of the channel between $x = 45$ and 205\,mm. The ``circle'' family consists of closed curves of four splines
through perturbed points. These outlines are rounded rather than circular, but the family keeps its name in the files.}
\end{figure}

\subsection*{Meshing}
Each geometry is written as a gmsh geometry file and meshed with gmsh 4.15.2\cite{gmsh}, using the ``Delaunay for
quads'' algorithm with recombination into a quad-dominant mesh. Every level has one uniform characteristic cell size
(2, 1, 0.5 or 0.25\,mm) and no local refinement (Design choices). The mesh is
extruded by one 10\,mm layer and converted with gmshToFoam. It is accepted when checkMesh passes all its default
checks (``Mesh OK''). A mesh that failed was generated again with additional smoothing (\texttt{Mesh.Smoothing = 5}).
This was needed for 3 meshes at 2\,mm, 3 in each 1\,mm run, 5 at 0.5\,mm and 14 at 0.25\,mm. The mesh variant is
recorded for every run. Meshing ran on one thread, so the meshes are deterministic.

\subsection*{Flow solution}
The steady incompressible Navier--Stokes equations are solved with simpleFoam of OpenFOAM v2512 (ESI)\cite{openfoam}
using the SIMPLEC algorithm\cite{simplec}. Convection is discretised with linearUpwind (second order) in all runs
except \run{base\_fo}, which uses first-order upwind. Each run is set up from the OpenFOAM case template of its
configuration. Gradients use the Gauss linear scheme and Laplacians the Gauss
linear corrected scheme, with one non-orthogonal corrector. Pressure is solved with GAMG and velocity with symmetric
Gauss--Seidel, and the velocity is under-relaxed with a factor of 0.9.

Every run is iterated until the initial residuals of $u_x$, $u_y$ and $p$ are all below $10^{-6}$. The initial
residual of a discretised equation $A\psi=b$ is computed at the start of an iteration, before its linear solve, with
the default normalisation of OpenFOAM v2512\cite{openfoam_norm}:
\begin{equation*}
r=\frac{\sum|b-A\psi|}{\sum\big(|A\psi-A\bar\psi|+|b-A\bar\psi|\big)},
\end{equation*}
where the sums run over the cells and $\bar\psi$ is the field's mean. This residual measures how far the fields are
from satisfying the discretised equations, rather than how much they change between iterations. In every template, the
linear solves within each iteration stop at $r<10^{-8}$ or once $r$ has fallen by a factor of 100 (pressure) or 10
(velocity). These solver tolerances do not decide when a run ends. The largest final residual in the dataset is
$9.99999\times10^{-7}$. A run that reached its iteration cap with its residuals still falling was continued from its
last time step until it converged. The longest runs needed 40,716 iterations at 2\,mm, 3,893 and 16,579 at 1\,mm
(first and second order), 15,647 at 0.5\,mm and 52,748 at 0.25\,mm. The iterations, cap and number of solver runs of
every run are in the metadata. At 2\,mm, 27 runs whose residuals had stalled between $10^{-3}$ and $3\times10^{-2}$
were instead started again from their converged 1\,mm solution. They used SIMPLE and under-relaxation (pressure 0.3,
velocity 0.7) in a template of their own, the relaxed template, but otherwise identical settings and discretisation.
These relaxed 2\,mm runs converged after 575--7,652 iterations.

Apart from the 27 relaxed 2\,mm runs, the 2\,mm and both 1\,mm runs start from rest. The 0.5 and 0.25\,mm runs start
from the converged 1\,mm second-order solution, mapped onto their mesh with mapFields. The effect of this start is
tested in Technical Validation.

\subsection*{Arrays}
The converged fields of every run are resampled to a regular grid with VTK 9.7.0\cite{vtk}. The reader
vtkOpenFOAMReader interpolates the cell values of $\mathbf{u}$ and $p$ to the mesh points. The filter
vtkResampleToImage then samples them on $n_x\times n_y$ points that cover the whole domain, both ends included, with
about one point per cell size (Table 2). ParaView's OpenFOAM reader and its Resample To Image filter are based on them.
Points inside the obstacle lie outside the mesh and are set to zero. Each sample has six channels (Table 3). Three are
the flow fields $u_x$, $u_y$ and $p$. The other three are geometry channels made from the same resampling: the
distance to the nearer channel wall, a cell-type mask and the signed distance to the obstacle outline. The wall
distance is $\min(j, n_y-1-j)$ for grid row $j$, so both wall rows have 0. The cell type is 1
on the two channel walls and at the obstacle points, 2 on the inlet column, 3 on the outlet column and 4 elsewhere.
The end points of the inlet and outlet columns take the inlet and outlet values, respectively. The signed distance is
the Euclidean distance, in grid indices, to the nearest obstacle point with at least one of its four neighbours
outside the obstacle. It is zero on these points and negative inside the obstacle. Because the grid spacings in $x$ and $y$ differ slightly
(2.011 and 2.034\,mm on the 2\,mm grid), it is not an exact physical distance. Fig.~3 shows one case in all five
configurations. At its scale, the 0.5 and 0.25\,mm fields are hard to tell apart by eye. Their attributes still
differ, for example by 0.64\,\% in the recirculation length (median, Technical Validation).

\begin{figure}[t]
\centering
\includegraphics[width=\textwidth]{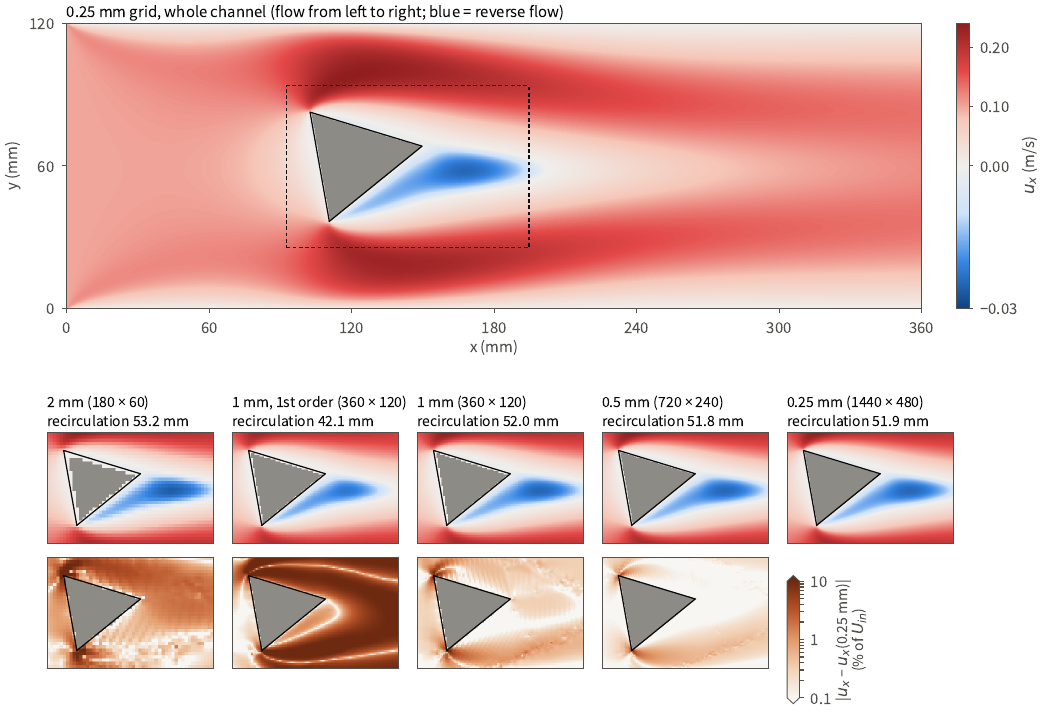}
\caption{\textbf{One case in all five configurations.} Top: streamwise velocity $u_x$ of a normal right triangle, a case
chosen at random, on the 0.25\,mm grid over the whole channel. Blue is reverse flow. Middle: the dashed window on every
grid, as stored (no smoothing), with the exact outline and the recirculation length computed on each mesh. Bottom:
$|u_x-u_x^{0.25}|$ on the grid of every cheaper configuration, in \% of $U_\mathrm{in}$ on a logarithmic scale. The
0.25\,mm array is interpolated bilinearly to the points of each of these grids, and points in the obstacle, or whose
interpolation touches it, are masked.}
\end{figure}

\begin{table}[t]
\centering\small
\caption{\textbf{Channels of every sample.} Points inside the obstacle are zero in channels 0--2.}
\begin{tabular}{@{}cllll@{}}
\toprule
Channel & Name & Content & Unit \\
\midrule
0 & \texttt{u\_x} & streamwise velocity & m\,s$^{-1}$ \\
1 & \texttt{u\_y} & cross-stream velocity & m\,s$^{-1}$ \\
2 & \texttt{p} & kinematic pressure, zero at the outlet & m$^2$\,s$^{-2}$ \\
3 & \texttt{wall\_distance} & distance to the nearer channel wall & grid points \\
4 & \texttt{cell\_type} & 1 wall or obstacle, 2 inlet, 3 outlet, 4 fluid & -- \\
5 & \texttt{obstacle\_sdf} & signed distance to the obstacle outline, negative inside & grid points \\
\bottomrule
\end{tabular}
\end{table}

\subsection*{Attributes}
The flow attributes of every run (Table 4) are computed on a copy of the converged case on the simulation mesh, where
the obstacle is a patch of its own. OpenFOAM function objects and utilities write the forces, the pressure and wall
shear on the obstacle faces, the cell fields and lines sampled across the channel. A Python script computes the
attributes from them\cite{code_pipeline}. The attributes, each defined in Table 4, fall into four groups, and the
outlet reverse flow checks the outlet condition. The forces on the obstacle are drag and lift, each with its pressure
and viscous (friction) parts. The second group is taken from the pressure and wall shear on the obstacle surface. The
pressure drop along the channel and the vorticity and dissipation describe the flow in the channel. The recirculation
zone, the flow split at the obstacle and a velocity profile behind it describe the shape of the wake. The
recirculation zone is the reverse flow ($u_x<0$) connected to the obstacle, so that a separate thin reverse-flow layer
along a channel wall does not count. The vorticity and the dissipation leave out the first 5\,mm of the channel, which
contain the singular inlet corners. Pressure and wall shear are also stored per obstacle face for every run.

The column \texttt{dissipation} is the pseudo-dissipation $\nu\int|\nabla\mathbf{u}|^2\,\mathrm{d}A$. It differs by
boundary terms from the viscous dissipation $2\nu\int\mathbf{S}\!:\!\mathbf{S}\,\mathrm{d}A$, where $\mathbf{S}$ is the
strain-rate tensor. In five runs checked, one of each configuration, the viscous dissipation over the region used here
was 0.16--0.51\,\% lower, and 0.16\,\% lower in the 0.25\,mm run. The outlet reverse flow is zero in every 0.25\,mm
run, so no reverse flow reaches the outlet, where the zero-gradient condition would otherwise prescribe the velocity
of inflowing fluid. The sensitivity of the solutions to the outlet position was not tested.

Three attributes are included but are not mesh-independent. The peak wall shear and the peak vorticity grow with
refinement at the sharp corners of the obstacles, where the exact solution is singular. The front pressure did not
converge with the mesh in two thirds of the cases, because the stagnation point often sits on a corner.

\subsection*{Label uncertainty}
Except for \texttt{outlet\_reverse\_flow}, which is zero in every 0.25\,mm run, each scalar attribute of each case is
computed at four second-order levels. The discretisation uncertainty $\U$ of its 0.25\,mm value $\phi_1$ is estimated
with the least-squares procedure of E\c{c}a and Hoekstra\cite{eca}. The procedure is designed for refinement data that
scatter and are not in the asymptotic range, as is the case for independent unstructured meshes. The step size of
level $i$ is computed from the cell counts as $h_i=(N_1/N_i)^{1/2}$, where $N_1$ is the number of cells of the
0.25\,mm mesh. A power law $\phi=\phi_0+a h^p$ with an observed order $p$, and laws with the fixed orders 1, 2 and
1+2, are fitted by least squares. Each is fitted both unweighted and with weights proportional to $1/h$, and the
weighting whose power-law fit has the smaller standard deviation $\sigma$ is used for all of them. The error estimate
is $e=|\phi_1-\phi_0|$, where $\phi_0$ is the extrapolated value of the fit that the observed order selects. For
$0.5\le p<2.1$, this is the power law, and the factor of safety is $F_s=1.25$. For $0<p<0.5$ or $p\ge2.1$, the
fixed-order fit with the smallest $\sigma$ is used instead, chosen from the orders 1, 2 and 1+2, or from 2 and 1+2
when $p\ge2.1$. Then $\phi_0$, $\sigma$ and the fitted curve $\phi_\mathrm{fit}$ are those of this fit, and $F_s=3$.
The uncertainty is $\U=F_s e+\sigma+|\phi_1-\phi_\mathrm{fit}(h_1)|$. When $\sigma$ is at least $\Delta\phi$, a third
of the range of the four values, the uncertainty is instead
$\U=(3\sigma/\Delta\phi)(e+\sigma+|\phi_1-\phi_\mathrm{fit}(h_1)|)$. Where the four values do not approach a limit in
a way the fits can describe (fitted order $p\le0$), $\U$ is three times the largest difference between them. This
typically happens because the scatter between the independent meshes exceeds the trend. Since all runs converged
iteratively, this behaviour concerns the convergence of an attribute with the mesh. The rule applies to 46\,\% of the
attribute--case pairs, including 42\,\% of the drag values and 93--96\,\% of the two peak quantities, but none of the
dissipation values. Where all four values are equal (217 pairs, all of them zero recirculation lengths and areas),
$\U=0$ by convention, although their numerical error need not be zero.

$\U$ estimates the discretisation error of the 0.25\,mm solution but does not bound it. For the three attributes that
are not mesh-independent (Attributes), the values are not shown to approach a finite limit, so their $\U$ is not an
uncertainty about a mesh-independent value. $\U$ does not include the loss in resampling to the arrays or the
iteration error, which was small against $\U$ in the cases tested in Technical Validation. The uncertainty file (Usage
Notes) gives $\U$ both in the units of the attribute and as a percentage of a scale. The scale is the value itself,
except for lift, its parts and the parts of the drag. These are taken relative to the drag, since the lift is small
and of either sign and the two drag parts shift in opposite directions with the mesh. The wake velocity is taken
relative to $U_\mathrm{in}$, and the front and base pressure relative to the dynamic pressure $\tfrac12
U_\mathrm{in}^2$. The percentage is undefined in the 227 pairs whose 0.25\,mm value is zero: 163 recirculation
lengths, which are zero when no reverse flow reaches behind the obstacle, and 64 recirculation areas. The level errors
in Technical Validation use the same scales, except for the recirculation length and area. Their scale is the larger
of the value and the obstacle height (its square for the area), so that short wakes do not give large relative values.
The scale differs from the value only in the cases whose zone is shorter than the obstacle height (or smaller in area
than its square). The same procedure was applied to the eight attributes that are also computed from the arrays, at
every level on its own grid, to measure how well the arrays represent them (Technical Validation). The three-level
grid convergence index\cite{roache} is released as well, where the convergence is monotonic. The uncertainty is
computed by \texttt{tools/uncertainty.py} and the grid convergence index by
\texttt{tools/mesh\_study.py}\cite{code_pipeline}.

\subsection*{Splits}
Three out-of-distribution (OOD) test sets, \texttt{ood\_blockage}, \texttt{ood\_rotation} and \texttt{ood\_position},
are held out so that models can be tested on a mild extrapolation. Each holds cases beyond the parameter range of the
training cases in one measure. The sets are tails of the distribution produced by the same generator, each formed from
the cases above the 95th percentile of its measure (Fig.~4). The measures are the blockage ratio (above 0.389), the
absolute rotation (above 41.8$^\circ$) and the distance of the position shift from the middle of its range, with
$\Delta x$ and $\Delta y$ each divided by their half range (above 1.196). The position set therefore holds out the
corners of the joint shift domain, and a shift that is extreme in one direction but central in the other can still be
in the training split. The three tails hold 224, 225 and 225 cases. Thirty cases lie in two tails (9 in blockage and
rotation, 8 in blockage and position, 13 in rotation and position) and none in all three. A case in more than one tail
goes to the first of these sets, so the sets hold 224, 216 and 204 cases. No case of the other splits lies in any of
the tails. The sets are not balanced over families and sizes. \texttt{ood\_blockage} holds mostly larger obstacles
(169 of 224) and triangles (159), and a single diamond. \texttt{ood\_rotation} holds mostly smaller and normal
obstacles (187 of 216). This is presumably because rotation raises the bounding-box height, so that strongly rotated
large obstacles more often exceed the blockage limit or fall in the blockage set first. The position set is close to
balanced. The other 3,856 cases are split 80/10/10 into training (3,085), validation (385) and test (386) cases within
every size class and family (Table 5). The cases of each size class and family are shuffled with a generator seeded
with the split seed (2026), the size and the family. The split depends only on the cases and is independent of their
order.

\begin{figure}[t]
\centering
\includegraphics[width=\textwidth]{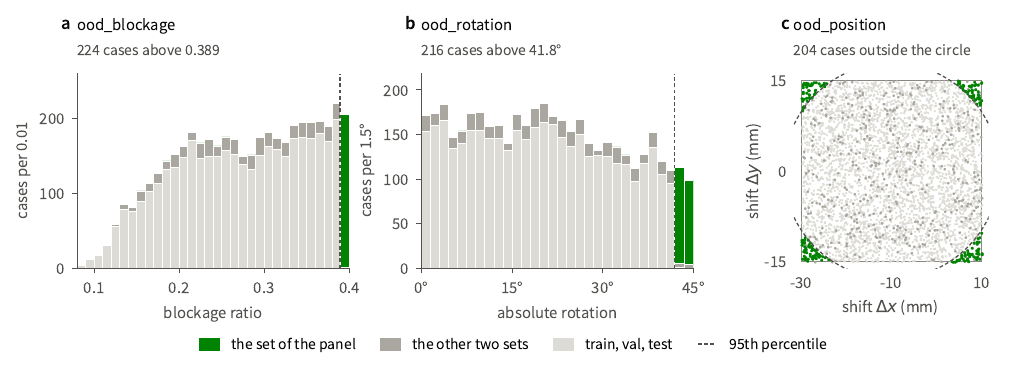}
\caption{\textbf{The out-of-distribution test sets.} Each set holds the cases above the 95th percentile (dashed) of
its measure. \textbf{a}, Blockage ratio. \textbf{b}, Absolute rotation. \textbf{c}, Distance of the position shift from
the middle of its range, with both shifts divided by their half ranges (outside the dashed circle). A case in more than
one tail belongs to the first of these sets.}
\end{figure}

\subsection*{Computation and provenance}
Every step that computed the data ran in one container image, built by continuous integration. The image contains
OpenFOAM v2512, gmsh 4.15.2, VTK 9.7.0 and Python 3.14 with numpy, scipy, h5py and pyarrow. All 22,500 runs used the
same image digest. Two steps that only unpack and read the run archives ran outside the image, because they need the
Python package zstandard, which the image lacks. The case of each run continued under a raised iteration cap was
restored from its archive on the cluster's login node, and the solver resumed inside the image. The wake velocity was
computed after the production, so its sampled lines were extracted from the archives in Google Colab.
The production ran on
the KUACC cluster of Ko\c{c} University, on Intel Xeon E5-2640 and E5-2670 v2 processors, with one SLURM job and one
CPU core per run. Peak solver memory was at most 1.05\,GB. The 22,500 runs took 35,922 solver core-hours, 94\,\% of
them at 0.25\,mm. A 0.25\,mm run costs about 185 times as much as a 1\,mm run (median). It has 16 times the cells,
needs 6 times the iterations and takes 1.9 times the time per iteration and cell on the same CPU model. The solver
time of a run is the simpleFoam execution time, summed over all solver runs when a run was continued over several
jobs. It does not include meshing, mapping or the prerequisite 1\,mm solution of a mapped run, and superseded attempts
are not counted. The task time covers all steps of the task that finished the run: set-up and meshing, mapping, the
solver, resampling, attributes, archiving and upload. Times are as measured on the two CPU models, without
normalisation. Every run records its convergence, set-up, cost and provenance (Data Records, Metadata). The complete
case of every run (mesh, the last two time steps, logs and attribute outputs) was archived, so any attribute can be
recomputed without solving again.

\subsection*{Use of large language models}
Claude (Anthropic), a large language model, was used to help write the analysis and pipeline code, search and
compare the literature, and draft and edit the manuscript. ChatGPT (OpenAI) was used for a final review of the text
for grammar and consistency. The authors checked all code, numbers and citations and
take full responsibility for the content.

\section*{Data Records}

The dataset is available on Hugging Face\cite{data_arrays} under the CC BY 4.0 licence. The release described here is
version 1.0.0. It consists of 38 data files and the checksum file \texttt{checksums.md5}, 39 files and 104.0\,GB in
all. Among the 38 files is a data card (\texttt{DATACARD.md}) with the caveats and provenance notes. The MD5 of
\texttt{checksums.md5} (\hash{82a46f931bb1d60b57dbc564305019fb}) identifies the release, and any change to the data would be
a new version. The archived OpenFOAM case of every run is on Hugging Face\cite{data_raw} (22,500 archives, 1.56\,TB).
The folder structure is:

{\small
\begin{verbatim}
v1.0.0/
  0.25x/ base_fo/ base_lu/ 4x/ 16x/   each with train.h5, val.h5, test.h5,
                                      ood_blockage.h5, ood_rotation.h5, ood_position.h5
  metadata.parquet  metadata.csv  mesh_study.parquet  surface.parquet
  label_uncertainty.parquet  dropped_cases.csv  dataset_info.json
  DATACARD.md  checksums.md5
\end{verbatim}
}

\textbf{Arrays.} Each file \texttt{<run>/<split>.h5} holds three HDF5 datasets. The array \texttt{fields} is float32 of
shape $(N, 6, n_x, n_y)$ and is stored with one sample per chunk and no compression, so a single sample can be read
without loading the file. The names in \texttt{case\_name} identify the case of each row, and \texttt{time} gives the
iteration of the exported solution. Row $n$ is the same case in every run of a split, in the order of the metadata, so
the configurations of a case are directly paired. The file attributes give the channel names (Table 3), the run, the grid, and
the mean and standard deviation of every channel over the training split of that run. The grid points include both
ends of the domain, so the spacing is $L/(n_x-1)$ and $H/(n_y-1)$. The axis $x$ points along the flow, and the inlet
is at index 0. The arrays of \run{0.25x}, \run{base\_fo}, \run{base\_lu}, \run{4x} and \run{16x} take 1.2, 4.7, 4.7,
18.7 and 74.7\,GB, and the tables add about 0.2\,GB.

\textbf{Metadata.} \texttt{metadata.parquet} (also as CSV) has one row per case and 259 columns. For each case, it gives
the split, the case name with its size, family and index, the global seed, the sampled width, height, rotation and
shift, the number of attempts and the outline points. The derived geometry comprises the bounding box, centroid,
area, blockage ratio and the gaps to the walls, inlet and outlet. For every run, columns named \texttt{<run>.<field>}
hold the 20 attributes of Table 4 and a record of the run. The record covers the convergence (status, iterations,
number of solver runs, final residuals), the mesh (mesh check, mesh variant, cell count) and the set-up (template,
starting fields, raised cap). It also gives the cost (solver and task time, peak memory) and the computing environment
(host, CPU model, allocated cores, SLURM job, container digest, code revision).

\begin{table}[tp]
\centering\footnotesize
\caption{\textbf{Attributes of every run}, computed by a Python script from the output of OpenFOAM function objects and
utilities on the simulation mesh (Methods, Attributes). Pressures and stresses are kinematic. Forces are for the
10\,mm depth of the mesh (multiply by 100 for values per metre) and equal newtons for $\rho = 1$\,kg\,m$^{-3}$.
Dissipation is per metre of depth.}
\begin{tabularx}{\textwidth}{@{}P{3.6cm}P{3.7cm}P{2.3cm}X@{}}
\toprule
Attribute & Column & Unit & Note \\
\midrule
drag: total, pressure, viscous & \texttt{Fx}, \texttt{Fx\_pressure}, \texttt{Fx\_viscous} & m$^4$\,s$^{-2}$ per 10\,mm &
\\
lift: total, pressure, viscous & \texttt{Fy}, \texttt{Fy\_pressure}, \texttt{Fy\_viscous} & m$^4$\,s$^{-2}$ per 10\,mm &
\\
front pressure & \texttt{front\_pressure} & m$^2$\,s$^{-2}$ & largest $p$ of the obstacle faces, not
mesh-independent (stagnation point often on a corner) \\
base pressure & \texttt{base\_pressure} & m$^2$\,s$^{-2}$ & unweighted mean $p$ of the faces in the rearmost
10\,\% of the obstacle's $x$ extent \\
mean wall shear & \texttt{mean\_wall\_shear} & m$^2$\,s$^{-2}$ & area-weighted $|\tau_w|/\rho$ on the obstacle \\
peak wall shear & \texttt{max\_wall\_shear} & m$^2$\,s$^{-2}$ & largest face value, not mesh-independent (singular at
corners) \\
pressure drop, inlet to outlet & \texttt{dp\_inlet\_outlet} & m$^2$\,s$^{-2}$ & difference of the area averages on
the inlet and outlet patches \\
pressure drop, cross-sections & \texttt{dp\_sections} & m$^2$\,s$^{-2}$ & the same on lines 10\,mm behind the inlet
and 10\,mm before the outlet \\
mean vorticity & \texttt{mean\_vorticity} & s$^{-1}$ & area-weighted mean $|\omega|$, first 5\,mm left out \\
peak vorticity & \texttt{peak\_vorticity} & s$^{-1}$ & largest cell $|\omega|$, first 5\,mm left out, not
mesh-independent (singular at corners) \\
pseudo-dissipation & \texttt{dissipation} & m$^4$\,s$^{-3}$ per metre & $\nu\int|\nabla\mathbf{u}|^2$, first
5\,mm left out \\
recirculation length & \texttt{recirculation\_length\_mm} & mm & furthest reverse-flow cell of the zone, from the
downstream end of the obstacle; zero if the zone does not reach behind it \\
recirculation area & \texttt{recirculation\_area\_mm2} & mm$^2$ & area of the reverse flow connected to the
obstacle \\
flow split & \texttt{flow\_split\_upper} & -- & flow rate above the obstacle over the total, with $u_x$ integrated
from each wall to the obstacle at the $x$ of its centroid \\
wake velocity & \texttt{wake\_min\_u\_x} & m\,s$^{-1}$ & smallest $u_x$ on a line across the channel 20\,mm behind
the rearmost outline point of the obstacle, within the obstacle's $y$ extent \\
outlet reverse flow & \texttt{outlet\_reverse\_flow} & -- & reverse flow rate in the last 5\,mm over the net flow
rate (area-weighted over cells), zero in every 0.25\,mm run \\
\bottomrule
\end{tabularx}
\end{table}

\textbf{Further tables.} \texttt{surface.parquet} holds the pressure and wall shear stress on every obstacle face for
every case and run, with the face position. \texttt{mesh\_study.parquet} has one row per case and attribute (85,500
rows). It covers 19 attributes, all except \texttt{outlet\_reverse\_flow}, which is zero in every 0.25\,mm run. Each
row holds a three-level study of the 1, 0.5 and 0.25\,mm levels. It gives the values of all five runs, the refinement
ratios, the type of convergence (monotonic, oscillatory, none or equal) and the observed order. Where the convergence
is monotonic, it also gives the extrapolated value, the error of every level against it and the grid convergence index
of the 0.25\,mm value. \texttt{label\_uncertainty.parquet} holds the E\c{c}a--Hoekstra uncertainty estimate of the 19
attributes of every case, also distributed with the benchmark repository\cite{code_bench}. \texttt{dropped\_cases.csv}
lists the parameters and final residuals of the 18 dropped cases, whose 1\,mm second-order run did not converge.
\texttt{dataset\_info.json} gives the counts per split, the runs and grids, the code revisions and the image digest.
\texttt{DATACARD.md} summarises how to read the files and gives the attributes that are not mesh-independent, the runs
that needed a raised cap or a different start, and the provenance notes.

\begin{table}[t]
\centering\small
\caption{\textbf{Splits.} Each split is one file per run, so every configuration has the same cases. ood:
out-of-distribution, the cases above the 95th percentile of the measure given (threshold in brackets). A case in more
than one tail is in the first of these sets.}
\begin{tabular}{@{}lrl@{}}
\toprule
Split & Cases & Content \\
\midrule
\texttt{train} & 3,085 & 80\,\% of the in-distribution cases of every size and family \\
\texttt{val} & 385 & 10\,\% \\
\texttt{test} & 386 & 10\,\% \\
\texttt{ood\_blockage} & 224 & blockage ratio (0.389) \\
\texttt{ood\_rotation} & 216 & absolute rotation (41.8$^\circ$) \\
\texttt{ood\_position} & 204 & distance of the shift from the middle of its range (1.196) \\
total & 4,500 & 300 per size and family \\
\bottomrule
\end{tabular}
\end{table}

\section*{Technical Validation}

\subsection*{Solver setup against a reference benchmark}
The steady flow past a cylinder in a channel at $Re = 20$ (DFG 2D-1\cite{dfg}) was solved with the discretisation,
SIMPLEC settings, $10^{-6}$ criterion and gmsh meshing of the production. Only the geometry, the viscosity and the
parabolic inlet were taken from the benchmark. Three uniform meshes were used (D/20, D/40 and D/80), and the two finer
runs started from the coarse solution, as in the production (Table 6). On the finest mesh, the drag coefficient and
the recirculation length lie inside the reference intervals. Between the meshes they change non-monotonically and by
less than the width of their intervals, so they were not extrapolated. The pressure difference is 0.8\,\% low on the
finest mesh, and its extrapolated value lies inside the interval. The lift coefficient is 9\,\% above the reference
value, a difference of 0.02\,\% of $c_D$, and its iteration error is 0.15\,\%. The lift is mostly viscous, and its
pressure part changes sign between D/40 and D/80, which points to the lack of wall refinement in the uniform meshes. No
wall-refined mesh was run to confirm this.

\begin{table}[t]
\centering\small
\caption{\textbf{DFG 2D-1 with the production settings.} $\Delta p$ in m$^2$\,s$^{-2}$, $L_a$ in m.}
\begin{tabular}{@{}lrrrrrr@{}}
\toprule
Mesh & Cells & Iterations & $c_D$ & $c_L$ & $\Delta p$ & $L_a$ \\
\midrule
D/20 & 35,965 & 410 & 5.5848 & $-0.0013$ & 0.1138 & 0.08428 \\
D/40 & 143,907 & 715 & 5.5857 & 0.0100 & 0.1157 & 0.08433 \\
D/80 & 575,745 & 1,977 & 5.5827 & 0.0116 & 0.1166 & 0.08446 \\
Extrapolated & & & -- & 0.0118 & 0.1175 & -- \\
Reference interval & & & 5.57--5.59 & 0.0104--0.0110 & 0.1172--0.1176 & 0.0842--0.0852 \\
\bottomrule
\end{tabular}
\end{table}

\subsection*{Iterative convergence and initial fields}
All 22,500 runs stopped with every initial residual below $10^{-6}$. The dataset audit (Completeness, integrity and
reproducibility) checked this from the provenance of every run. The median number of iterations rises from 209 at
2\,mm to 3,497 at 0.25\,mm, with a long tail (Fig.~5). The figure shows the computational effort. To test whether the
criterion is strict enough, 12 converged 0.25\,mm runs were continued from their released state with a criterion of
$10^{-8}$. They were the four runs that had needed the most iterations (41,173--52,748) and three runs at the blockage
limit of 0.40. The other five were the four with the smallest uncertainty of drag or dissipation and one with the
median iteration count. Ten of them reached $10^{-8}$ after 1,790 to 28,097 further iterations. The two slowest were
stopped after about 30,000 further iterations, with largest residuals of $6\times10^{-8}$ and $8\times10^{-8}$ that
were still falling. The fields changed by 0.03--0.05\,\% (RMS, largest in $u_y$). Each attribute changed by at most
0.21 of the label uncertainty of its case, with a median of 0.0002 over the 228 attribute--case pairs. Only three
pairs exceed 0.1, all in one of the cases chosen for their small uncertainty. In the cases tested, the iteration error
left at $10^{-6}$ is small against the discretisation uncertainty.

In production, the 0.5 and 0.25\,mm runs start from the mapped 1\,mm solution, as do the 27 relaxed 2\,mm runs
(Methods, Flow solution). Cases were solved on the same mesh both from rest and from these starting fields (Table 7). The remaining
differences grow as the mesh gets finer and are of the order of the iteration error left at the $10^{-6}$ stop. At
0.25\,mm, the largest attribute difference of each case is 0.3--8\,\% of its label uncertainty. The mapped starting
fields reduced the number of iterations.

\begin{figure}[t]
\centering
\includegraphics[width=0.62\textwidth]{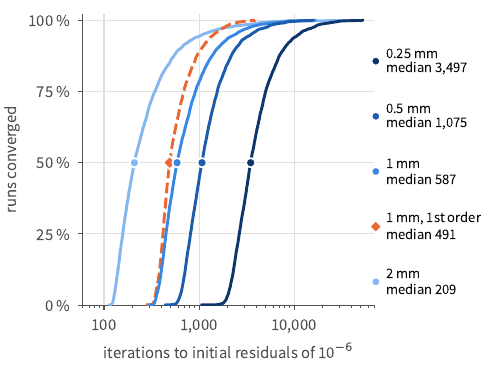}
\caption{\textbf{Iterations to initial residuals of $10^{-6}$.} Share of the 4,500 runs of each configuration that had
converged within a given number of iterations. The dot on each curve marks the median, and the first-order 1\,mm run
is dashed.}
\end{figure}

\begin{table}[t]
\centering\small
\caption{\textbf{Same mesh, different starting fields.} Largest difference over the cases, between runs from rest
and from the mapped 1\,mm solution. Fields: RMS of the difference over the RMS of the field.}
\begin{tabular}{@{}lclll@{}}
\toprule
Level & Cases & Fields & Attributes & Iterations, mapped vs from rest \\
\midrule
2\,mm (relaxed runs) & 4 & 0.007\,\% ($u_y$) & 0.001\,\% (base pressure) & 575--6,157 vs 6,909--14,576 \\
0.5\,mm & 6 & 0.021\,\% ($u_y$) & 0.033\,\% (base pressure) & 760--2,385 vs 1,217--4,782 \\
0.25\,mm & 6 & 0.10\,\% ($u_y$) & 0.15\,\% (base pressure) & 2,466--8,142 vs 3,966--15,885 \\
\bottomrule
\end{tabular}
\end{table}

\subsection*{A looser stopping criterion}
To quantify how the labels of fine meshes depend on the stopping criterion, 113 cases were solved again with
initial-residual targets of $10^{-3}$ (velocity) and $10^{-2}$ (pressure) and otherwise identical settings. They are
105 test cases, drawn from seven bins of the iteration count of their converged 0.25\,mm run, and 8 cases of the
blockage set. Table 8 and Fig.~6 report the test cases, weighted by the inverse inclusion probability. Started from rest, the
loose 0.25\,mm run stops after a tenth of the iterations that the converged run needs, and its recirculation length
is about a third too short. Started from the 1\,mm solution, it stops almost at once, and the label stays close to
the coarse solution. The coarse-to-fine difference is then understated, at 0.36\,\% instead of 0.47\,\%
(relative RMS of $u_x$, Fig.~6a). With both levels loose from rest, the difference
is overstated by a factor of 36 (17.1\,\%). The loose criterion barely affects the drag and recirculation length of the
coarse 1\,mm run. Because the direction of
the error depends on the starting fields, these results do not support a single correction applied afterwards to all runs.

\begin{table}[t]
\centering\footnotesize
\caption{\textbf{Loose against converged runs.} Weighted medians over the 105 test cases, with the 90th percentile of
$|\cdot|$ in brackets. Iterations are given at the loose stop and for the converged run, and the RMS change of $u_x$ in \% of
$U_\mathrm{in}$.}
\begin{tabular}{@{}lrrrrr@{}}
\toprule
Run & Loose stop & Converged & RMS change of $u_x$ & Drag & Recirculation length \\
\midrule
1\,mm, from rest & 233 & 571 & 0.55 & $+0.01$\,\% (0.0) & $0.00$\,\% (0.1) \\
0.25\,mm, from 1\,mm & 8 & 3,353 & 0.32 & $+1.33$\,\% (2.3) & $-1.53$\,\% (2.9) \\
0.25\,mm, from rest & 355 & 3,353 & 18.67 & $+0.50$\,\% (7.8) & $-32.6$\,\% (67.2) \\
\bottomrule
\end{tabular}
\end{table}

\begin{figure}[t]
\centering
\includegraphics[width=\textwidth]{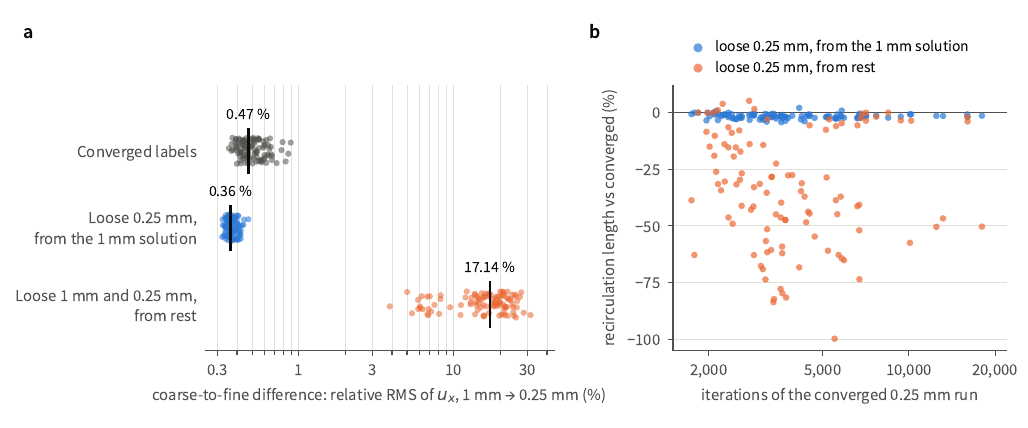}
\caption{\textbf{Effect of a loose stopping criterion on the labels.} \textbf{a}, Coarse-to-fine difference of every
test case (relative RMS of $u_x$, 1\,mm interpolated bicubically to 0.25\,mm after the zeros inside the obstacle are
replaced by the nearest fluid value). It is shown for the converged labels,
for loose 0.25\,mm labels started from the 1\,mm solution and for both levels loose from rest. Ticks mark the weighted
medians. \textbf{b}, Change of the recirculation length of the loose 0.25\,mm labels, started from the 1\,mm solution
and from rest, against the number of iterations their converged run needed.}
\end{figure}

\subsection*{Discretisation error and label uncertainty}
Table 9 and Fig.~7 compare the error of each cheaper configuration with the E\c{c}a--Hoekstra uncertainty of the 0.25\,mm
label, on the scales of the level errors (Methods, Label uncertainty). The second-order levels converge towards
the label. Their median error falls with every refinement for every attribute except the mean wall shear, and by half
or more from 1 to 0.5\,mm. For the drag it is 1.10, 0.76 and 0.33\,\% at 2, 1 and 0.5\,mm, and for the dissipation
4.68, 1.69 and 0.53\,\%. The mean wall shear error rises from 2 to 1\,mm (1.14 and 1.48\,\%) before it falls to
1.00\,\% at 0.5\,mm. The first-order 1\,mm run differs most from the label where the flow separates. Its recirculation
length is 11\,\% off and its drag 3.9\,\%, whereas its pressure drag is close to the label (0.97\,\%). Its
recirculation length is shorter than the label in 96\,\% of the cases, which is consistent with the numerical diffusion of
first-order upwinding.

Because medians do not describe single cases, Fig.~7b gives the share of cases whose error exceeds their own
uncertainty estimate. Where the estimate is zero, any difference counts as exceeding it. At 0.5\,mm this share is
below 9\,\% for every attribute except dissipation (61\,\%). For the drag and the recirculation length it is 20 and
29\,\% at 1\,mm, 42 and 52\,\% at 2\,mm, and 85 and 90\,\% for the first-order run. The dissipation error exceeds its
uncertainty in nearly every case at 1 and 2\,mm, and the mean-vorticity error does so in half the cases at 1\,mm. For
the first-order run, the share is lowest for the pressure and friction drag (6 and 32\,\%). For the second-order
levels, these shares are not an independent test, because $\U$ is estimated from the same four levels. They describe
where each level lies relative to the estimated error of the label.

The meshes of a case are independent unstructured meshes, not nested refinements, so part of the discretisation error
is irregular. The three-level grid convergence index is undefined where the convergence is not monotonic. This is the
case for 34\,\% of the drag labels, 31--38\,\% of its parts, 45\,\% of the recirculation lengths and 43\,\% of the
wake velocities. The E\c{c}a--Hoekstra procedure handles such data with least-squares fits and larger factors of
safety. Where the data converge cleanly, as for dissipation, it gives a tighter uncertainty than the simple estimate
$3|\phi_{0.25}-\phi_{0.5}|$ (0.49 against 1.59\,\%). The friction and pressure drag converge slowly on their own
(median observed orders 0.29 and 0.36), but relative to the drag their errors at 0.5\,mm are about 1\,\% or less
(Table 9). The three attributes that are not mesh-independent often do not converge with the mesh. For example, the
front pressure did not converge in 2,947 of the 4,500 cases.

\begin{table}[t]
\centering\footnotesize
\caption{\textbf{Error of every configuration against the 0.25\,mm label, and the uncertainty $\U$ of the label}, in
\%. Median over the 4,500 cases, with the 90th percentile in brackets. Thirteen of the 20 attributes are shown, and
the uncertainty of all 19 is in the uncertainty file. Errors and $\U$ are on the scales of the level errors (Methods,
Label uncertainty).}
\begin{tabular}{@{}lrrrrr@{}}
\toprule
Attribute & 2\,mm & 1\,mm, first order & 1\,mm & 0.5\,mm & $\U$ of the label \\
 & \run{0.25x} & \run{base\_fo} & \run{base\_lu} & \run{4x} & \run{16x} \\
\midrule
drag & 1.10 (1.82) & 3.93 (5.17) & 0.76 (1.25) & 0.33 (0.56) & 1.45 (3.79) \\
pressure drag & 2.72 (4.54) & 0.97 (1.83) & 2.18 (3.38) & 1.02 (1.59) & 5.98 (11.35) \\
friction drag & 1.95 (3.41) & 2.89 (4.22) & 1.46 (2.45) & 0.69 (1.19) & 4.07 (8.70) \\
lift & 0.66 (1.93) & 1.34 (3.40) & 0.46 (1.11) & 0.23 (0.50) & 0.97 (2.97) \\
recirculation length & 2.80 (6.30) & 11.10 (17.38) & 1.62 (3.45) & 0.64 (1.40) & 2.41 (8.05) \\
recirculation area & 1.75 (3.80) & 7.07 (13.22) & 1.10 (2.33) & 0.49 (1.04) & 2.19 (6.90) \\
wake velocity & 0.63 (1.45) & 2.79 (5.06) & 0.41 (0.78) & 0.18 (0.33) & 0.73 (2.05) \\
base pressure & 1.20 (3.12) & 2.99 (8.43) & 0.58 (1.70) & 0.29 (0.94) & 1.50 (5.44) \\
dissipation & 4.68 (5.67) & 4.56 (7.07) & 1.69 (2.38) & 0.53 (0.80) & 0.49 (1.04) \\
mean vorticity & 0.65 (0.94) & 2.31 (3.63) & 0.33 (0.52) & 0.12 (0.20) & 0.34 (0.98) \\
$\Delta p$ between sections & 0.58 (1.02) & 2.33 (3.00) & 0.47 (0.76) & 0.20 (0.35) & 1.00 (2.35) \\
flow split & 0.17 (0.54) & 0.29 (0.85) & 0.09 (0.25) & 0.04 (0.12) & 0.20 (0.68) \\
mean wall shear & 1.14 (3.33) & 4.90 (7.75) & 1.48 (3.28) & 1.00 (2.11) & 4.58 (10.53) \\
\bottomrule
\end{tabular}
\end{table}

\begin{figure}[t]
\centering
\includegraphics[width=\textwidth]{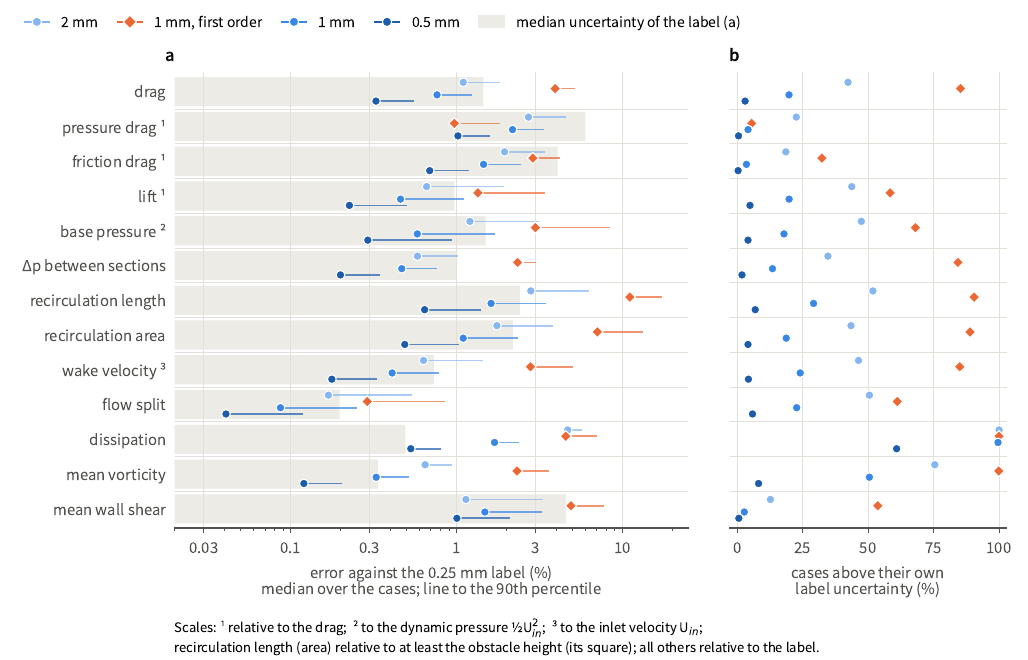}
\caption{\textbf{Error of every cheaper configuration against the 0.25\,mm label, and the label uncertainty.} \textbf{a},
Dots mark the median over the 4,500 cases, and lines extend to the 90th percentile of the cases (not a confidence
interval). The grey band extends to the median uncertainty of the label, which is not a limit for single cases.
\textbf{b}, Share of cases whose error exceeds their own uncertainty estimate. Scales as in Table 9.}
\end{figure}

\subsection*{Resampled arrays against the simulation mesh}
The dataset audit checked every array of every run for its shape and finite values, zero values inside the obstacle,
geometry channels consistent with the outline, and mass conservation. The obstacle area agreed with the sampled
geometry to within 0.67\,\% at 2\,mm and 0.02\,\% at 0.25\,mm (median). For mass conservation, the volume flux through
cross-sections agreed with the inlet flux to within 0.69\,\% at 2\,mm and 0.022\,\% at 0.25\,mm. To measure what the
resampling loses, the eight attributes that can be computed from the fields were also computed from the arrays of
every configuration, each on its own grid. They were compared with the values on the simulation mesh (Table 10).
Gradient quantities lose the most. For dissipation and mean vorticity, the loss is first order in the grid spacing and
halves with every refinement, from 13 to 1.7\,\% for dissipation. The arrays systematically underestimate both, which
is consistent with the cell-to-point interpolation smoothing the steep gradients at the walls. The inlet-to-outlet
pressure drop of the arrays is about 1.4\,\% low at every level. The array value averages the end columns of the grid,
whereas the mesh value averages the patch faces, which include the singular inlet corners. Peak vorticity, taken as
the largest grid value, is 22--36\,\% low. It is not mesh-independent on the simulation mesh either.

The uncertainty procedure was also applied to the array values of the four levels, each on its own grid. The median
uncertainty of the 0.25\,mm array values is then 2.94\,\% for dissipation (0.49\,\% for the mesh values) and
0.90\,\% for mean vorticity (0.34\,\%). For the recirculation length and area, the flow split and the pressure drop,
it is about the same as on the mesh. For gradient quantities, the stored arrays represent the flow less accurately
than the attributes computed on the mesh.

\begin{table}[t]
\centering\small
\caption{\textbf{Attributes from the arrays against the simulation mesh.} Median $|\text{array}-\text{mesh}|$ over the
4,500 cases in \% of a scale, with the median signed difference in brackets. The scale is the mesh value of the same
run, except for the recirculation length and area, relative to at least the obstacle height and its square, and the
wake velocity, relative to $U_\mathrm{in}$.}
\begin{tabular}{@{}lrrrrr@{}}
\toprule
Attribute & 2\,mm & 1\,mm, first order & 1\,mm & 0.5\,mm & 0.25\,mm \\
 & \run{0.25x} & \run{base\_fo} & \run{base\_lu} & \run{4x} & \run{16x} \\
\midrule
recirculation length & 2.00 ($+0.87$) & 1.09 ($+0.67$) & 1.05 ($+0.69$) & 0.52 ($+0.39$) & 0.25 ($+0.18$) \\
recirculation area & 2.51 ($-2.51$) & 0.76 ($-0.76$) & 0.80 ($-0.80$) & 0.25 ($-0.24$) & 0.07 ($-0.07$) \\
flow split & 0.09 ($+0.01$) & 0.02 (0.00) & 0.02 (0.00) & 0.01 (0.00) & 0.00 (0.00) \\
$\Delta p$ inlet to outlet & 1.52 ($-1.52$) & 1.41 ($-1.41$) & 1.46 ($-1.46$) & 1.40 ($-1.40$) & 1.36 ($-1.36$) \\
dissipation & 13.01 ($-13.01$) & 7.17 ($-7.17$) & 6.80 ($-6.80$) & 3.44 ($-3.44$) & 1.74 ($-1.74$) \\
mean vorticity & 4.35 ($-4.35$) & 2.25 ($-2.25$) & 2.14 ($-2.14$) & 1.05 ($-1.05$) & 0.52 ($-0.52$) \\
peak vorticity & 35.7 ($-35.7$) & 31.3 ($-31.3$) & 29.2 ($-29.2$) & 24.6 ($-24.6$) & 21.5 ($-21.5$) \\
wake velocity & 0.63 ($-0.22$) & 0.43 ($-0.37$) & 0.34 ($-0.22$) & 0.19 ($-0.14$) & 0.09 ($-0.08$) \\
\bottomrule
\end{tabular}
\end{table}

\subsection*{Completeness, integrity and reproducibility}
Before the release, an audit ran over every run of every case, and all its checks passed. It checked the membership
(300 cases per size and family, five runs each, the dropped cases and the reserves used) and the convergence and
provenance of every run. It also checked the archives (one per run, with the size and MD5 of all 22,500 archives on
the remote store) and the arrays and attributes. To detect mixed-up cases, every configuration was compared with 0.25\,mm,
and no case had outliers in three or more attributes. The audit covered the mesh study and the splits, and found every
geometry file regenerated from the seed identical. Its only warnings concern eight 0.25\,mm runs whose starting fields
were not recorded in their provenance. Their archives show that they started from the 1\,mm solution like all the
others, and the metadata records the starting fields of every run. In the release files, the splits equal those of the
audit, row $n$ is the same case in every run, and all 38 checksums are correct. Five archives, one per configuration,
were restored and processed again with the code at tag \texttt{data-v1.0.0}. They include a run continued over four
solver runs, a 0.5\,mm run continued under a raised cap and one of the relaxed 2\,mm runs. The arrays and attributes
obtained from them were identical bit for bit to the released ones. With the same code, three cases were solved again
from rest at 2\,mm and at 1\,mm (second order). They needed the same iterations and gave identical arrays and
attributes.

\section*{Usage Notes}

With h5py, \verb|h5py.File("16x/test.h5")["fields"][n]| returns the $(6, 1440, 480)$ sample of row $n$ without
loading the rest of the file. The rows follow the order of the split's cases in the metadata (column
\texttt{dataset\_index}), so pairs of configurations need no matching beyond the row index. Every file also stores the
\texttt{case\_name} of each row. Grid point $(i, j)$ lies at $x = iL/(n_x-1)$ and $y = jH/(n_y-1)$ in mm, with the
inlet at $i = 0$. The grids of different configurations share the end points but not the interior points, so two of
them cannot be compared without interpolation.

The value 1 of \texttt{cell\_type} marks both the obstacle and the two channel walls. A mask of the fluid is
\texttt{cell\_type != 1}, and \texttt{obstacle\_sdf <= 0} marks the obstacle alone. The signed distance is zero on the
outermost obstacle points, so \texttt{obstacle\_sdf < 0} leaves them out. Because the inlet corners are singular,
error measures of gradients are best taken without the first 5\,mm of the channel, as for the vorticity and
dissipation attributes. The normalisation statistics in each file (Data Records) are computed over all grid points,
including the zero values inside the obstacle. Statistics of the test
or out-of-distribution sets should not be used for normalisation. The cell type is a category and should not be
normalised with its statistics.

The attributes in the metadata are the values on the simulation mesh, and the same quantity computed from the arrays
differs from them by the resampling loss (Table 10). Values from predicted fields should therefore be compared with
values computed in the same way from the label arrays. Such a comparison measures agreement with the stored arrays
rather than accuracy against the solution on the mesh, because the two resampling losses need not cancel. The
uncertainty of the 19 attributes of every case is in \texttt{label\_uncertainty.parquet} (Data Records). Rows with
\texttt{source = openfoam} are the uncertainty of the 0.25\,mm solution, and rows with \texttt{source = array}
describe the convergence of the array values. The column \texttt{method} gives the basis of each estimate:
\texttt{power}, \texttt{low\_order} or \texttt{high\_order} for a fitted estimate, \texttt{anomalous} for the fallback
of three times the largest difference, and \texttt{equal} where all four values are equal.

For every run, the metadata records the starting fields, any raised iteration cap, the number of solver runs, the mesh
variant and the template, so runs that deviate in any of these respects can be found or left out. Each run records the
code revision deployed when it finished. For a run continued over several jobs, this revision also holds for the
earlier jobs, because the container image and the attribute, resampling and array code did not change during
production. The recorded revision does not hold for the wake velocity and the wall distance (channel 3). Both were
computed for every run after the production, and the code at tag \texttt{data-v1.0.0} (builder revision
\texttt{a21e54b} in \texttt{dataset\_info.json}) gives the released values. Every geometry can be regenerated from the
seed with the pipeline\cite{code_pipeline}. The archived cases
are complete OpenFOAM cases, which can be opened with the pipeline's container image to compute further quantities or
other resamplings.

\section*{Data Availability}
\sloppy
The arrays, tables, data card and checksums of version 1.0.0 (Data Records) are available on
Hugging Face\cite{data_arrays}. The archived OpenFOAM case of every run is available on Hugging Face\cite{data_raw}. The
label uncertainty is also distributed with the benchmark repository\cite{code_bench}.

\fussy

\section*{Code Availability}
The pipeline that generated the data\cite{code_pipeline}, tag \texttt{data-v1.0.0}, is available under the MIT
licence. It covers geometry sampling, meshing, the OpenFOAM case templates, resampling and the attributes, as well as
the production on a SLURM cluster and the dataset audit. Every step that computed the data ran in one container image,
\texttt{ghcr.io/asaydam/ms-cofd-2dsl} (digest \hash{sha256:ae5bc11fb90901b2d7cab7012fb91a211a268a865f19aef7cd2b3d6e6a2d0364}), with OpenFOAM v2512 (ESI), gmsh 4.15.2, VTK 9.7.0 and
Python 3.14. Two steps that only unpack and read the run archives ran outside it (Methods, Computation and
provenance). The settings are in \texttt{config/dataset.yaml} and the case templates. The per-run exceptions (the
raised caps and the relaxed template) are recorded in the metadata, and every run records its code revision. The label
uncertainty is computed by \texttt{tools/uncertainty.py}. Tables 9 and 10 are computed by
\texttt{validation/level\_errors.py} and \texttt{validation/array\_vs\_openfoam.py}, respectively, and the other
checks of Technical Validation by the scripts in \texttt{validation/}, documented in \texttt{docs/validation.md}. The
benchmark repository\cite{code_bench} holds readers, interpolation baselines and the label-uncertainty file.

\section*{Author Contributions}
A.S. and B.A. designed the dataset. A.S. developed the data-generation pipeline and the benchmark code, carried out
the simulations and the technical validation, and wrote the manuscript. B.A. proposed the force-based quantities,
supervised the project, and reviewed and edited the manuscript.

\section*{Competing Interests}
The authors declare no competing interests.

\section*{Funding}
This work is supported by the KUIS AI Center of Ko\c{c} University, Istanbul. We gratefully acknowledge this support.

\end{document}